\documentstyle[twoside,epsf,12pt]{article}

\catcode`\@=11
\long\def\@makefntext#1{
\protect\noindent \hbox to 3.2pt {\hskip-.9pt
$^{{\eightrm\@thefnmark}}$\hfil}#1\hfill}               

\def\@makefnmark{\hbox to 0pt{$^{\@thefnmark}$\hss}}    

\def\ps@myheadings{\let\@mkboth\@gobbletwo
\def\@oddhead{\hbox{}
\rightmark\hfil\eightrm\thepage}
\def\@oddfoot{}\def\@evenhead{\eightrm\thepage\hfil
\leftmark\hbox{}}\def\@evenfoot{}
\def\sectionmark##1{}\def\subsectionmark##1{}}

\oddsidemargin=\evensidemargin
\newcounter{sectionc}\newcounter{subsectionc}\newcounter{subsubsectionc}
\renewcommand{\section}[1] {\vspace{12pt}\addtocounter{sectionc}{1}
\setcounter{subsectionc}{0}\setcounter{subsubsectionc}{0}\noindent
        {\tenbf\thesectionc. #1}\par\vspace{5pt}}
\renewcommand{\subsection}[1] {\vspace{12pt}\addtocounter{subsectionc}{1}
        \setcounter{subsubsectionc}{0}\noindent
{\bf\thesectionc.\thesubsectionc. {\kern1pt \bfit #1}}\par\vspace{5pt}}
\renewcommand{\subsubsection}[1] {\vspace{12pt}
\addtocounter{subsubsectionc}{1}
        \noindent{\tenrm\thesectionc.\thesubsectionc.\thesubsubsectionc.
        {\kern1pt \tenit #1}}\par\vspace{5pt}}
\newcommand{\nonumsection}[1] {\vspace{12pt}\noindent{\tenbf #1}
        \par\vspace{5pt}}

\newcounter{appendixc}
\newcounter{subappendixc}[appendixc]
\newcounter{subsubappendixc}[subappendixc]
\renewcommand{\thesubappendixc}{\Alph{appendixc}.\arabic{subappendixc}}
\renewcommand{\thesubsubappendixc}
        {\Alph{appendixc}.\arabic{subappendixc}.\arabic{subsubappendixc}}

\renewcommand{\appendix}[1] {\vspace{12pt}
        \refstepcounter{appendixc}
        \setcounter{figure}{0}
        \setcounter{table}{0}
        \setcounter{lemma}{0}
        \setcounter{theorem}{0}
        \setcounter{corollary}{0}
        \setcounter{definition}{0}
        \setcounter{equation}{0}
     \renewcommand{\thefigure}{\Alph{appendixc}.\arabic{figure}}
     \renewcommand{\thetable}{\Alph{appendixc}.\arabic{table}}
     \renewcommand{\theappendixc}{\Alph{appendixc}}
     \renewcommand{\thelemma}{\Alph{appendixc}.\arabic{lemma}}
     \renewcommand{\thetheorem}{\Alph{appendixc}.\arabic{theorem}}
     \renewcommand{\thedefinition}{\Alph{appendixc}.\arabic{definition}}
        \renewcommand{\thecorollary}{\Alph{appendixc}.\arabic{corollary}}
        \renewcommand{\theequation}{\Alph{appendixc}.\arabic{equation}}
        \noindent{\tenbf Appendix \theappendixc #1}\par\vspace{5pt}}
\newcommand{\subappendix}[1] {\vspace{12pt}
        \refstepcounter{subappendixc}
        \noindent{\bf Appendix \thesubappendixc. {\kern1pt \bfit #1}}
        \par\vspace{5pt}}
\newcommand{\subsubappendix}[1] {\vspace{12pt}
        \refstepcounter{subsubappendixc}
        \noindent{\rm Appendix \thesubsubappendixc. {\kern1pt \tenit #1}}
        \par\vspace{5pt}}

\newcommand{\textlineskip}{\baselineskip=13pt}
\newcommand{\smalllineskip}{\baselineskip=10pt}

\def\eightcirc{
\begin{picture}(0,0)
\put(4.4,1.8){\circle{6.5}}
\end{picture}}
\def\eightcopyright{\eightcirc\kern2.7pt\hbox{\eightrm c}}

\newcommand{\copyrightheading}[1]
        {\vspace*{-2.5cm}\smalllineskip{\flushleft
        {\footnotesize $\eightcopyright$\, World Scientific Publishing
         Company}\\
         }}

\def\abstracts#1#2#3{{
        \centering{\begin{minipage}{4.5in}\baselineskip=10pt\footnotesize
        \parindent=0pt #1\par
        \parindent=15pt #2\par
        \parindent=15pt #3
        \end{minipage}}\par}}

\renewenvironment{thebibliography}[1]
        {\frenchspacing
         \ninerm\baselineskip=11pt
         \begin{list}{\arabic{enumi}.}
        {\usecounter{enumi}\setlength{\parsep}{0pt}
         \setlength{\leftmargin 12.7pt}{\rightmargin 0pt} 
         \setlength{\itemsep}{0pt} \settowidth
        {\labelwidth}{#1.}\sloppy}}{\end{list}}

\newcounter{itemlistc}
\newcounter{romanlistc}
\newcounter{alphlistc}
\newcounter{arabiclistc}

\newcommand{\fcaption}[1]{
        \refstepcounter{figure}
        \setbox\@tempboxa = \hbox{\footnotesize Fig.~\thefigure. #1}
        \ifdim \wd\@tempboxa > 5in
           {\begin{center}
        \parbox{5in}{\footnotesize\smalllineskip Fig.~\thefigure. #1}
            \end{center}}
        \else
             {\begin{center}
             {\footnotesize Fig.~\thefigure. #1}
              \end{center}}
        \fi}

\newcommand{\tcaption}[1]{
        \refstepcounter{table}
        \setbox\@tempboxa = \hbox{\footnotesize Table~\thetable. #1}
        \ifdim \wd\@tempboxa > 5in
           {\begin{center}
        \parbox{5in}{\footnotesize\smalllineskip Table~\thetable. #1}
            \end{center}}
        \else
             {\begin{center}
             {\footnotesize Table~\thetable. #1}
              \end{center}}
        \fi}

\def\@citex[#1]#2{\if@filesw\immediate\write\@auxout
        {\string\citation{#2}}\fi
\def\@citea{}\@cite{\@for\@citeb:=#2\do
        {\@citea\def\@citea{,}\@ifundefined
        {b@\@citeb}{{\bf ?}\@warning
        {Citation `\@citeb' on page \thepage \space undefined}}
        {\csname b@\@citeb\endcsname}}}{#1}}

\newif\if@cghi
\def\cite{\@cghitrue\@ifnextchar [{\@tempswatrue
        \@citex}{\@tempswafalse\@citex[]}}
\def\citelow{\@cghifalse\@ifnextchar [{\@tempswatrue
        \@citex}{\@tempswafalse\@citex[]}}
\def\@cite#1#2{{$\null^{#1}$\if@tempswa\typeout
        {IJCGA warning: optional citation argument
        ignored: `#2'} \fi}}

\def\pmb#1{\setbox0=\hbox{#1}
        \kern-.025em\copy0\kern-\wd0
        \kern.05em\copy0\kern-\wd0
        \kern-.025em\raise.0433em\box0}

\def\fnt#1#2{\footnotetext{\kern-.3em
        {$^{\mbox{\scriptsize #1}}$}{#2}}}

\def\fpage#1{\begingroup
\voffset=.3in
\thispagestyle{empty}\begin{table}[b]\centerline{\footnotesize #1}
        \end{table}\endgroup}

\def\runninghead#1#2{\pagestyle{myheadings}
\markboth{{\protect\footnotesize\it{\quad #1}}\hfill}
{\hfill{\protect\footnotesize\it{#2\quad}}}}
\font\tenrm=cmr10
\font\tenit=cmti10
\font\tenbf=cmbx10
\font\bfit=cmbxti10 at 10pt
\font\ninerm=cmr9

\font\eightrm=cmr8

\def\qed{\hbox{${\vcenter{\vbox{                        
   \hrule height 0.4pt\hbox{\vrule width 0.4pt height 6pt
   \kern5pt\vrule width 0.4pt}\hrule height 0.4pt}}}$}}

\def\bsc{{\sc a\kern-6.4pt\sc a\kern-6.4pt\sc a}}       
\def\bflatex{\bf L\kern-.30em\raise.3ex\hbox{\bsc}\kern-.14em
T\kern-.1667em\lower.7ex\hbox{E}\kern-.125em X}

\begin{document}

\runninghead{Numerical Approach to Two-Loop Integrals with Masses}
{Numerical Approach to Two-Loop Integrals with Masses}
\footnotesize
\begin{flushright}
KEK CP-028 \\
KEK Preprint 95-28
\end{flushright}

\normalsize\textlineskip
\thispagestyle{empty}
\setcounter{page}{1}

\copyrightheading{}                

\vspace*{0.88truein}

\fpage{1}
\centerline{\bf NUMERICAL APPROACH TO }
\vspace*{0.045truein}
\centerline{\bf TWO-LOOP THREE POINT FUNCTIONS WITH MASSES\footnote{
Presented by K. Kato at the AI-HENP 95 workshop, Pisa, April 1995}}
\vspace*{0.35truein}
\centerline{\footnotesize Junpei FUJIMOTO\footnote{junpei@minami.kek.jp}
, Yoshimitsu SHIMIZU\footnote{shimiz@minami.kek.jp} }
\vspace*{0.020truein}
\centerline{\footnotesize\it
National Laboratory for High-Energy Physics (KEK)}
\baselineskip=10pt
\centerline{\footnotesize\it Oho 1-1, Tsukuba, Ibaraki 305, Japan}
\vspace*{8pt}
\centerline{\footnotesize Kiyoshi KATO\footnote{kato@cc.kogakuin.ac.jp}}
\vspace*{0.020truein}
\centerline{\footnotesize\it Kogakuin University}
\baselineskip=10pt
\centerline{\footnotesize\it Nishi-Shinjuku 1-24,
Shinjuku, Tokyo 160, Japan}
\vspace*{8pt}
\centerline{\normalsize and}
\vspace*{8pt}
\centerline{\footnotesize Toshiaki
KANEKO\footnote{
On leave of absence from
Meiji-Gakuin University , Totsuka, Yokohama 244, Japan,\\
kaneko@minami.kek.jp }}
\vspace*{0.020truein}
\centerline{\footnotesize\it Laboratoire d'Annecy-le-Vieux de Physique
des Particules}
\vspace*{0.015truein}
\centerline{\footnotesize\it (LAPP), B.P. 110,F-74941 Annecy-le Vieux
 Cedex, FRANCE}


\vspace*{0.21truein}
\abstracts{
Extending the method successful for one-loop integrals,
the computation of two-loop diagrams with general internal
masses is discussed.
For the two-loop vertex of non-planar type,
as an example, we show a calculation related to
\(Z^0\rightarrow t \bar t\) vertex.
}{}{}



\vspace*{1pt}\textlineskip      

\vspace*{10pt}

\newcommand{\footn}[1]%
  {\footnote{#1}\hspace{2mm}}
\newcommand{\refeq}[1]{Eq.(\ref{eq:#1})}
\newcommand{\reffig}[1]{Fig.\ref{fig:#1}}

The calculation of loop integrals is essential
to obtain the precise theoretical prediction for
high-energy reactions.
Loop integrals for one-loop diagrams
can be expressed by logarithms and dilogarithms\cite{thv}.
For a class of higher order diagrams, compact analytic
expressions are obtained.
For instance, some two-loop two point functions with single mass(\(m\))
are given by functions\cite{broadhurst}
of \(x=s/m^2\).
As two-loop diagrams in the electro-weak theory
include complicated mass combinations,
an analytic formula seems not available.
In general, for the case with more than two independent
dimensionless variables,
it seems to be impossible to obtain a compact
analytic formula by polylogarithms and so forth.
In scattering processes, we have two or more invariants and in
the electroweak theory we encounter diagrams with many
different masses.
Thus the theoretical prediction in the electroweak theory
requires  a method to compute the loop integrals
for arbitrary mass scales. To meet this request,
we have already developed numerical methods for  two-loop integrals.
There exist a few other approaches for
this direction by momentum space integral,\cite{kleimer}
momentum expansion with Pad\'e approximation,\cite{tarasov}
asymptotic expansion\cite{davydychov}, and so forth.

Automatic generation of two-loop diagrams is possible
in the standard model\cite{kaneko}.
{}From the generated diagrams,
symbolic expression for the numerator and the loop integrand
can be generated. The latter, the integrand in two-loop as
a function of Feynman parameter, is treated numerically

Our strategy is as follows\cite{ourwork}:
\begin{itemize}
 \item Any loop integral can be given by the definite integral
in Feynman parameter space.
 \item The integrand can be singular where the denominator becomes zero.
 If there is no singularity, the value of integral can be obtained
without any trouble.  The numerical accuracy is only limited by
available computer time.
 \item If there exists singularity,
 the numerical integration is done after either of the following:
 \begin{enumerate}
 \item (Symmetric method) The integrand is replaced by a symmetric sum
 of the function so as to eliminate the singularity.
 \item (Hybrid method) The integral for a few of variables is
 done analytically until the singularity is replaced by
 an integrable function.
 \end{enumerate}
\end{itemize}
In this report, we present results for those integrals appearing in the
non-planar vertex functions.

We consider the following scalar integral for vertex function
\begin{equation}
I=\int \frac{d^n l_1}{(2\pi)^n} \frac{d^n l_2}{(2\pi)^n}
\frac{1}{D_1D_2D_3D_4D_5D_6}
   \label{eq:twlp01}
\end{equation}
where \(1/D_j\)'s are propagator functions.
Introduction of Feynman parameters transforms it in the following form.
\begin{equation}
 I= \frac{1}{(4\pi)^n} \int_0^1 dx_1\cdots dx_6 \delta(1-\sum x)
    \frac{1}{U^{n/2}(V-i\epsilon)^{6-n}}
   \label{eq:twlp02}
\end{equation}
where \(U\) and \(V\) can be determined from
the topology of the diagram \cite{kinoshita}.
If we confine ourselves to a class of diagrams
where there is no ultraviolet divergence in a subgraph,
we can take \(n=4\).
The function \(U\) is
\begin{equation}
  U = \sum_{\bar{T}} \prod_{x_j\in \bar{T}} x_j
\end{equation}
where the summation is taken over {\it co-trees}(\(\bar{T}\))
in the diagram.  The function \(V\) is
given by
\begin{equation}
  V = \sum_j x_j m_j^2 - \frac{1}{U} \sum_S W_S p_S^2
\end{equation}
where the second sum is taken over {\it cutsets}(\(S\)) in the diagram.

External momenta, \(p_1, p_2, p_3\) are defined to flow
{\it inward} into the vertex and they satisfy
\( p_1+p_2+p_3=0\).
We use the notation
\(s_i = p_i^2 \),
\(x_i+x_j+\cdots=x_{ij\cdots}\), and
\(\bar x_{i\cdots} = 1-x_{i\cdots} \).
In the following, we consider the integral
\begin{equation}
  J = \int_0^1 dx_1\cdots dx_6 \delta(1-\sum x)
  \frac{1}{({\cal D}+i \epsilon)^2}
\end{equation}
where
\begin{equation}
  {\cal D}=\sum_S W_S p_S^2-U \sum_j x_j m_j^2
\end{equation}
and
\begin{equation}
  \sum W_S p_S^2 = f_1 s_1 + f_2 s_2 + f_3 s_3 \ .
\end{equation}

\begin{figure}[hbt]
\setlength{\unitlength}{1mm}
\begin{picture}(140,35)
  \put(10,30){{\bf (a)}}

  \put(35,30){\vector(0,-1){5}}
  \put(10,0){\vector(1,1){5}}
  \put(60,0){\vector(-1,1){5}}

  \put(15, 5){\vector(1,1){5}}
  \put(20,10){\vector(1,1){5}}
  \put(25,15){\vector(1,1){5}}
  \put(30,20){\vector(1,1){5}}

  \put(35,25){\vector(1,-1){5}}
  \put(40,20){\vector(1,-1){5}}
  \put(45,15){\vector(1,-1){5}}
  \put(50,10){\vector(1,-1){5}}

  \put(45,15){\vector(-1,0){10}}
  \put(35,15){\vector(-1,0){10}}
  \put(55, 5){\vector(-1,0){20}}
  \put(35, 5){\vector(-1,0){20}}

  \put(25,20){\(x_1 \)}
  \put(15,10){\(x_4 \)}
  \put(41,20){\(x_2 \)}
  \put(51,10){\(x_5 \)}
  \put(34,16){\(x_3 \)}
  \put(34,6){\(x_6 \)}

  \put(36,30){\(p_3\)}
  \put(5,1){\(p_1 \)}
  \put(61,1){\(p_2 \)}

  \put(80,30){{\bf (b)}}

  \put(105,30){\vector(0,-1){5}}
  \put(80,0){\vector(1,1){5}}
  \put(130,0){\vector(-1,1){5}}

  \put(85, 5){\vector(1,1){5}}
  \put(90,10){\vector(1,1){5}}
  \put(95,15){\vector(1,1){5}}
  \put(100,20){\vector(1,1){5}}

  \put(105,25){\vector(1,-1){5}}
  \put(110,20){\vector(1,-1){5}}
  \put(115,15){\vector(1,-1){5}}
  \put(120,10){\vector(1,-1){5}}

  \put(115,15){\vector(-3,-1){15}}
  \put(100,10){\vector(-3,-1){15}}
  \put(125, 5){\vector(-3,1){15}}
  \put(110,10){\vector(-3,1){15}}

  \put(95,20){\(x_1 \)}
  \put(85,10){\(x_4 \)}
  \put(111,20){\(x_2 \)}
  \put(121,10){\(x_5 \)}
  \put(94,6){\(x_3 \)}
  \put(114,6){\(x_6 \)}

  \put(106,30){\(p_3\)}
  \put(75,1){\(p_1 \)}
  \put(131,1){\(p_2 \)}

\end{picture}
\caption{Two-loop vertex diagrams. (a)planar type.  (b) non-planar type.}
  \label{fig:twlp1}
\end{figure}
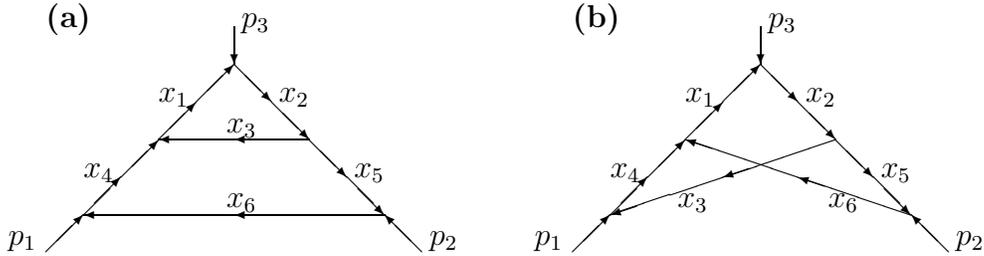

For the diagrams in \reffig{twlp1},
we obtain the following results.


\noindent
(a) Planar type
\begin{equation}
  U = x_{12}x_{3456}+x_3x_{456}
\end{equation}

\begin{equation}
  \begin{array}{c}
  f_1 = x_{123}x_4x_6+x_1x_3x_5 \\
  f_2 = x_{123}x_5x_6+x_2x_3x_4 \\
  f_3 = x_1x_2x_{3456}+x_{123}x_4x_5+x_1x_3x_5+x_2x_3x_4
  \end{array}
\end{equation}

\noindent
(b) Non-planar type
\begin{equation}
  U = x_{12}x_{3456}+x_{34}x_{56}
\end{equation}

\begin{equation}
  \begin{array}{c}
  f_1 = x_{1256}x_3x_4+x_1x_3x_6+x_2x_4x_5 \\
  f_2 = x_{1234}x_5x_6+x_2x_3x_6+x_1x_4x_5 \\
  f_3 = x_{3456}x_1x_2+x_2x_4x_6+x_1x_3x_5
  \end{array}
\end{equation}

For the first case, we have already reported the numerical
results\cite{ourwork}.
Here, we report the results for the non-planar vertex.

First, variables are transformed as follows.
\begin{equation}
  \begin{array}{cc}
  x_1=z_3(1+y_3)/2, & x_2=z_3(1-y_3)/2 \\
  x_3=z_1(1-y_1)/2, & x_4=z_1(1+y_1)/2 \\
  x_5=z_2(1-y_2)/2, & x_6=z_2(1+y_2)/2 \\
  \end{array}
\end{equation}
After the transform, three-fold symmetry in the non-planar
vertex can be seen clearly.
The integral becomes
\begin{equation}
  J = \frac{1}{8}\int_0^1 dz_1 dz_2 dz_3 \delta(1-\sum z) \hat z
     \int_{-1}^1 dy_1 \int_{-1}^1 dy_2 \int_{-1}^1 dy_3
    \frac{1}{({\cal D}+i \epsilon)^2}
\end{equation}
where \(\hat z=z_1z_2z_3\),
and the function in the denominator becomes
\begin{equation}
{\cal D} = {}^t\vec y A \vec y + \vec b\cdot \vec y +c
\label{eq:denomi}
\end{equation}
where
\begin{equation}
\vec y= \left(
  \begin{array}{c} y_1 \\ y_2 \\ y_3
 \end{array}
\right),
\end{equation}

\begin{equation}
A= \frac{1}{4}\left(
  \begin{array}{ccc}
-z_1^2\bar z_1s_1 & \hat z (-s_1-s_2+s_3)/2 & \hat z (-s_1+s_2-s_3)/2\\
\hat z (-s_1-s_2+s_3)/2 &-z_2^2\bar z_2s_2  & \hat z ( s_1-s_2-s_3)/2\\
\hat z (-s_1+s_2-s_3)/2 & \hat z (s_1-s_2-s_3)/2  & -z_3^2\bar z_3s_3
 \end{array}
\right),
\end{equation}
\begin{equation}\vec b= \frac{1}{2}\left(
  \begin{array}{c}
  z_1 (-m_3^2+m_4^2) \\
  z_2 (-m_5^2+m_6^2) \\
  z_3 (m_1^2-m_2^2)
 \end{array}
\right),
\end{equation}
\begin{equation}
c = \frac{1}{4}U\left[ z_1s_1+z_2s_2+z_3s_3
  -2(m_3^2+m_4^2)z_1  -2(m_5^2+m_6^2)z_2  -2(m_1^2+m_2^2)z_3\right] ,
\end{equation}
\begin{equation}
 U=z_1z_2+z_2z_3+z_3z_1 .
\end{equation}

So the problem is transformed into how to carry out the double integral
of a function similar to that which appears in  a box diagram
in one-loop.
The determinant of \(A\) is given by
\begin{equation}
{\rm det}A=\frac{1}{4^4}\hat z^2U(z_1s_1+z_2s_2+z_3s_3)
(s_1^2+s_2^2+s_3^2-2s_1s_2-2s_2s_3-2s_1s_3).
\end{equation}

As an example, we try to calculate ``\(Z_0\) exchange
for \(t\bar t\) vertex'' i.e.,
\begin{equation}
  \begin{array}{cc}
  p_1^2=p_2^2=m^2, & p_3^2=s, \\
  m_1=m_2=m_4=m_5=m,& m_3=m_6=M, \\
  m=150{\rm (GeV)},& M=91.17{\rm (GeV)}.
  \end{array}
 \label{eq:sample}
\end{equation}

Below the threshold, \(s<4m^2\), the integral has no
singularity and it can be done easily by adaptive Monte-Carlo
integration program BASES.\cite{bases}
Above threshold, the singularity, \({\cal D}=0\) , appears.
It is hyperboloid of one sheet or that of two sheets in \(\vec y\)
space.  The transition of the topology
occurs at \(c'=0\) where \(c'\) is defined by
\begin{equation}
{\cal D} = {}^t\vec y' A \vec y'  +c,' \qquad
c'=c-\frac{1}{4}{}^t\vec b A^{-1} \vec b.
\end{equation}
For the case in \refeq{sample}, the condition \(c'=0\) becomes
\begin{equation}
z_3=t, \qquad (M^2-s)(s+3M^2-4m^2)t^2+2M^2(s-M^2)t+M^2(4m^2-M^2)=0.
\end{equation}

By use of these formulas, we can calculate the value of integral
based on the hybrid method
as is described in detail in references.\cite{ourwork,ouroneloop}

\begin{figure}[hbt]
\setlength{\unitlength}{1mm}
\begin{picture}(140,35)
  \put(10,30){{\footnotesize Two-body cut}}

  \put(35,30){\vector(0,-1){5}}
  \put(10,0){\vector(1,1){5}}
  \put(60,0){\vector(-1,1){5}}

  \put(15, 5){\vector(1,1){5}}
  \put(20,10){\vector(1,1){5}}
  \put(25,15){\vector(1,1){5}}
  \put(30,20){\vector(1,1){5}}

  \put(35,25){\vector(1,-1){5}}
  \put(40,20){\vector(1,-1){5}}
  \put(45,15){\vector(1,-1){5}}
  \put(50,10){\vector(1,-1){5}}

  \put(45,15){\vector(-3,-1){15}}
  \put(30,10){\vector(-3,-1){15}}
  \put(55, 5){\vector(-3,1){15}}
  \put(40,10){\vector(-3,1){15}}

  \put(80,30){{\footnotesize Three-body cut}}

  \put(105,30){\vector(0,-1){5}}
  \put(80,0){\vector(1,1){5}}
  \put(130,0){\vector(-1,1){5}}

  \put(85, 5){\vector(1,1){5}}
  \put(90,10){\vector(1,1){5}}
  \put(95,15){\vector(1,1){5}}
  \put(100,20){\vector(1,1){5}}

  \put(105,25){\vector(1,-1){5}}
  \put(110,20){\vector(1,-1){5}}
  \put(115,15){\vector(1,-1){5}}
  \put(120,10){\vector(1,-1){5}}

  \put(115,15){\vector(-3,-1){15}}
  \put(100,10){\vector(-3,-1){15}}
  \put(125, 5){\vector(-3,1){15}}
  \put(110,10){\vector(-3,1){15}}

\thicklines

\multiput(23,20)(5,0){5}{\line(1,0){3}}
\multiput(93,23)(6,-3){6}{\line(2,-1){4}}

\end{picture}
  \caption{Cuts for non-planar vertex.}
  \label{fig:twlp2}
\end{figure}
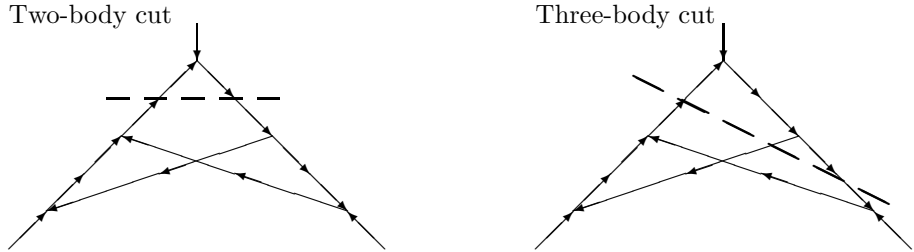
The calculation of the non-planar vertex at hand can be done
easily by the dispersion integral.
With respect to the variable \(s\), we have a two-body cut
as in \reffig{twlp2}(a), and three-body cuts as in
\reffig{twlp2}(b) and the reversed one.
\begin{equation}
  J(s)=\frac{1}{\pi}\int \frac{ \Im T(s')}{s'-s-i\epsilon} ds'
  ,\qquad
\Im T(s') = \Im T_{two}+ \Im T_{three}
\label{eq:disp}
\end{equation}
Here
\begin{equation}
\Im T_{two}= \int T_0(p_3 \rightarrow k_1+k_2) d\Gamma_2(k_1,k_2)
             T_{box}(k_1+k_2 \rightarrow p_1+p_2)
 \label{eq:disptwo}
\end{equation}
and
\begin{equation}
\Im T_{three}= \int T_0(p_3 \rightarrow k_1+k_3+k_5)
d\Gamma_3(k_1,k_3,k_5)
             T_0(k_1+k_3+k_5 \rightarrow p_1+p_2)+{\rm (reversed)}.
 \label{eq:dispthree}
\end{equation}
where tree amplitudes, \(T_0\) are expressed just by
product of propagator(s), and \(d\Gamma_n\) stands for the \(n\)-body
phase space. The one-loop box integral in \refeq{disptwo} is
\begin{equation}
T_{box}= \int dxdydz
\frac{1}{(-xyt-z(1-x-y-z)u+(x+y)M^2+(1-x-y)^2m^2)^2}
 \label{eq:dispbox}
\end{equation}
and is free from singularity because the denominator
of is non-negative.
Hence the 4-dimensional
integrals in \refeq{disptwo} and \refeq{dispthree}
are well behaved.
The singularity at \(s'=s\) in \refeq{disp} can be
handled by casting the integral into the form
\begin{equation}
  \Re J(s)=\frac{1}{2\pi}\int
  \frac{ \Im T(s+\sigma)- \Im T(s-\sigma)}{\sigma} d\sigma.
\label{eq:disp1}
\end{equation}
By use of BASES, this is evaluated as 5-dimensional
integral together all the integral variables, Feynman parameters
and kinematical ones.

\begin{figure}[htb]
\setlength{\unitlength}{1mm}
  \caption{Real part of non-planar vertex.}
  \label{fig:twlp3}
\end{figure}

In \reffig{twlp3}, we present the final results by these methods.
The calculation of the integrals in Feynman parameter space
is done at the same \(s\) repeatedly by changing the assignment
of external momenta and masses in \refeq{sample}.
This diagnostics test works well for the
points in the figures.

In this report, we presented only one example for
the real calculation.  However, the methods given here
work  in principle  for any mass parameters and
external momenta\footn{One needs more sophisticated treatment
if the infrared divergence exists.}.
The inclusion of numerator and the extension to four-point function
are interesting and will be studied in the coming research.

\nonumsection{Acknowledgements}
\noindent
This work is supported in part by the Ministry of Education,
Science and Culture, Japan under the
Grant-in-Aid for International Scientific Research Program No.04044158,
and the Special Research Fund of the Kogakuin University.

\nonumsection{References}

\end{document}